# Cancellation of the collisional frequency shift in caesium fountain clocks

K. Szymaniec[1], W. Chałupczak[1], E. Tiesinga[2], C.J. Williams[2], S. Weyers[3], R. Wynands[3]

[1] *National Physical Laboratory, Hampton Road, Teddington, TW11 0LW, UK*

[2] *Joint Quantum Institute and Atomic Physic Division, National Institute of Standards and Technology, 100 Bureau Drive, Gaithersburg, Maryland 20899-8423, USA*

[3] *Physikalisch-Technische Bundesanstalt, Bundesallee 100, 38116 Braunschweig, Germany*

We have observed that the collisional frequency shift in primary caesium fountain clocks varies with the clock state population composition and, in particular, is zero for a given fraction of the $|F = 4, m_F = 0\rangle$ atoms, depending on the initial cloud parameters. We present a theoretical model explaining our observations. The possibility of the collisional shift cancellation implies an improvement in the performance of caesium fountain standards and a simplification in their operation. Our results also have implications for test operation of fountains at multiple $\pi/2$ pulse areas.

The primary caesium clocks have increased their accuracy and stability by an order of magnitude over the last decade [1]. This staggering progress has mainly resulted from the implementations of laser cooling and the fountain configuration. Slow atoms allow for long interrogation times and observation of narrow resonances ($\leq 1$ Hz) leading to improvements in the short-term stability. At the same time many systematic effects, which limited the performance of thermal beam clocks, have been significantly reduced in the fountains resulting in improvements of their accuracy. The use of ultracold atoms gave rise, however, to a frequency shift due to collisions, an effect generally neglected in thermal beam clocks. For the several operational fountain frequency standards, the collisions constitute one of the major systematic effects limiting the standards' performance.

Generally, the fountain standards are corrected for the collisional frequency shift by extrapolating the measured frequency to zero atom density, assuming a linear dependence between the shift and the density. However, the atomic density is not measured directly in the fountains. The changes in the density are derived from changes in the number $N_{at}$ of detected atoms. A potential deviation from linearity between $N_{at}$ and the atomic density (and hence the shift) gives rise to an uncertainty of the correction. One way to minimize this uncertainty is to operate the fountain at a very low density [2]. Unfortunately, with a low atom number the detection signal-to-noise ratio (and short-term stability of the standard) is reduced. Another way relies on the implementation of a technique based on adiabatic passage to link unambiguously changes in density with changes in $N_{at}$ [3]. In this case, the accuracy of the collisional shift measurement is ultimately limited by a residual amount of atoms in $|F = 3, m_F \neq 0\rangle$ colliding with atoms in the clock states during the ballistic flight [4].

Ways of cancelling the collisional shift have been studied earlier. The cancellation would be possible if the cross-sections of the frequency changing collisions of the two clock states $|3,0\rangle$ and $|4,0\rangle$ were of opposite sign. Then, if a certain composition of the clock states were excited the two contributions to the clock shift would cancel. In the 1990s the relevant cross-sections for various Cs isotopes were calculated and it was shown that the cancellation is possible for radioactive isotopes [135]Cs and [137]Cs, but not for [133]Cs, which is used to realise the SI second [5]. Also in an earlier experiment no significant difference between the collisional shift for balanced and imbalanced populations of the clock states was found [6]. The calculations in Ref. [5] were based on a theoretical model developed in Ref. [7], predicting the Wigner threshold regime in Cs for microkelvin temperatures. More recently, as more accurate constraints for the Cs-Cs scattering parameters were given by experiments [8], new calculations were performed [9]. It was found that the Wigner threshold regime is reached only for temperatures (collision energies) below 0.1 nK. Thus, unlike previously assumed, the collision rate (and clock shift) does depend on the energy of the colliding atoms. The variation of the collisional shift with the clock state composition was predicted for temperatures below 0.2 $\mu$K.

In this paper, we present updated theoretical calculations of the collision rate coefficients and show that the sub-microkelvin temperatures are effectively reached in caesium fountains, if the atoms are launched at a $\mu$K temperature and in a small initial cloud ($e^{-1/2}$ radius $\sigma_0$, where typically $\sigma_0 < 1$ mm), e.g. when released from a magneto-optical trap (MOT). We then report on an experimental demonstration, in two independent systems, of the strong variation of the collisional shift as a function of the clock state composition and the cancellation of the shift for a particular ratio of the populations. We also discuss the implications of this effect for the operation and performance of caesium fountain clocks.

We have performed a calculation of the clock shift, the collision-induced shift in the transition frequency between the two clock states, based on the description of Refs. [10, 11]. In essence, the energy levels of an atom are shifted due to collisions or interactions with other atoms. The differential level shift of the two clock states is the clock shift. The atom-atom interaction, to good approximation, is determined by the interplay between two Born-Oppenheimer potentials, which describe the electronic bonding between ground-state caesium atoms, the rotation of the di-atom, and the hyperfine interaction of each atom. As the depth of the Born-Oppenheimer potentials is orders of magnitude larger than the collision energies and the spatial extent of the Born-Oppenheimer potentials is much smaller than the mean distance between the caesium atoms the clock shift is expressible in terms of scattering amplitudes or $S$ matrices that depend on relative collision energy $E$, relative orbital angular momentum, as well as the hyperfine states of the colliding atoms. Only in the Wigner threshold regime can a simplified as well as insightful expression be used [12].

Within the theory of [10, 11] the clock shift due to collisions at a single collision energy $E$ is given by $\Delta\omega(E) = n\sum_i \lambda_i(E)\rho_i$; where the sum is over the atomic hyperfine ground states, $\rho_i$ is the fraction of atoms in state $i$, and $n$ is the total atom number density. The collision rate coefficients $\lambda_i(E)$ are determined by (products of) scattering amplitudes. We assume that in the experiment only the states $i = |3,0\rangle$ and $|4,0\rangle$ are populated ($\rho_{30} + \rho_{40} = 1$).

Figure 1 shows our theoretical results for the collision rate coefficients $\lambda_{30}(E)$ and $\lambda_{40}(E)$ as a function of collision energy $E$. The results are based on the Born-Oppenheimer potentials of Ref. [13], and a close-coupled calculation of the scattering amplitudes. Most notable is that the energy dependence of the two rate coefficients is very different. The rate coefficient $\lambda_{30}(E)$ changes sign at $E/k_B = 0.16$ $\mu$K while $\lambda_{40}(E)$ is always negative. By controlling the atom fractions (or state composition) the collision energy at which $\Delta\omega(E)$ changes sign can be controlled.

In the experiment we have a distribution of collision energies, which evolves during the ballistic flight. We must modify $\Delta\omega(E)$ to include an average over the spatial and velocity density $n(\mathbf{x},\mathbf{v},t)$ of the spreading atom cloud and to find the instantaneous clock shift: $\Delta\omega(t) = \frac{1}{N} \sum_i \left\langle \lambda_i\left(\frac{\mu}{2}[\mathbf{v}_1 - \mathbf{v}_2]^2\right) n(\mathbf{x},\mathbf{v}_1,t) n(\mathbf{x},\mathbf{v}_2,t) \right\rangle \rho_i$, where $\langle \rangle$ denotes integration over any spatial and velocity coordinate, $\mu$ is the reduced mass of two colliding atoms, $N = \langle n(\mathbf{x},\mathbf{v},t) \rangle$ is the number of colliding atoms, and time $t$ satisfies $t_{up} < t < t_{down}$ ($t_{up}$ and $t_{down}$ are the instants of passage through the Ramsey cavity, respectively, and $t = 0$ is the launch time). Our observed clock shift is the instantaneous clock shift $\Delta\omega(t)$ averaged over the time the atoms spend between the two microwave interactions at $t_{up}$ and $t_{down}$ (Ramsey time). We calculated the clock shift by performing 3D Monte Carlo simulations of the expanding cloud [14].

The collision energies during the Ramsey time may be significantly lower than those derived from the launch temperature, typically 1-2 $\mu$K. The low collision energies result from a purely kinetic effect, position-velocity correlations, which develop in the expanding atomic cloud [15, 16]. As the cloud expands, collisions are only possible for the atoms with small relative velocities $v_{coll} < \sigma_0/t$. In order to gain insight into the evolution of the collision energies, we computed a local collision energy, defined as an average kinetic energy of atoms in a neighbourhood of a given atom in its reference frame. The local energy value does not depend on the size of the chosen neighbourhood, as long as it is smaller than the initial cloud size. Fig. 2 shows the evolution of the local energy for various launch temperatures $T_0$ and two initial cloud sizes $\sigma_0$. The initial sizes $\sigma_0 = 5$ mm and $\sigma_0 = 1$ mm correspond to situations when atoms are collected in an optical molasses or in a MOT, respectively. While the local energy decreases in both cases, the decrease is almost an order of magnitude larger for the small initial cloud. We further consider only small initial clouds, which are relevant to our experimental conditions. We note that, as the atomic density in the expanding cloud also decreases rapidly during the flight in the fountain, the collisions taking place during the first part of the Ramsey time have the prevailing effect on the frequency shift [14, 16].

For $T_0 = 1$ $\mu$K, $\sigma_0 < 1$ mm and $t > t_{up}$ the collision energies divided by the Boltzmann constant are less than 300 nK. For these collision energies the rate coefficients $\lambda_{30}$ and $\lambda_{40}$ differ in sign (Fig. 1), which gives rise to a strong variation of the collisional clock shift if the composition of the clock states (after the first Ramsey interaction) is varied. In the inset to Fig. 3 we show the variable shift calculated for parameters similar to those met in experiment (assuming spherical symmetry of the atom cloud). Of particular interest is the composition for which the collisional shift is zero ($\rho_{40}^{(C)}$). We note that the collisional shift cancellation occurs for a relatively narrow range of initial cloud sizes (Fig. 3). These sizes coincide with the typical values of $\sigma_0$ for those fountains where the atoms are released from a MOT. We also note that in the experiment reported in [6] the MOT contained a very large number of atoms ($10^9$), making the initial cloud larger ($\sigma_0 = 1.7$ mm) and hotter ($T_0 = 2.8$ $\mu$K), thus the shift variation was not observed and no cancellation could occur.

Our experiments were performed on two entirely independent primary frequency standards: NPL-CsF1 at National Physical Laboratory and PTB-CSF1 at Physikalisch-Technische Bundesanstalt. The set-ups, already described in detail earlier [17, 18], operate in cyclic modes: $10^7$-$10^8$ Cs atoms from a room temperature vapour are collected and cooled in a MOT. The atoms are then launched in moving optical molasses and further cooled. They typically reach an apogee at 85 cm (NPL) or 83 cm (PTB), which is, respectively, 31 cm and 39 cm above the centre of the Ramsey cavity. The launch temperature is found by measuring the time-of-flight width of the expanded cloud. The size of the cloud at launch is measured by imaging its fluorescence pattern on a CCD camera. The atoms are launched and prepared in the $|F = 3, m_F = 0\rangle$ state by microwave and optical pulses. After the atoms have completed their flight in the fountain, passing twice through the Ramsey microwave cavity, the populations of $|F = 3\rangle$ and $|F = 4\rangle$ are detected by laser-induced fluorescence. With two separate laser beams and two detectors the atomic populations can be measured and normalized to the total population.

The collisional shift is found by operating the fountain at two different atom numbers (densities) and comparing the respective frequencies measured against a reference frequency standard (hydrogen maser). The linearity between $N_{at}$ and the collisional frequency shift (i.e. average atomic density in the fountain) has been evaluated experimentally [17, 18]. The atom number is for instance set by varying the amplitude of the preparing microwave pulse.

The collisional shift is expected to vary linearly with $\rho_{40}$ (Fig. 3), which can be changed conveniently by adjusting the amplitude of the microwave field in the Ramsey cavity: $\rho_{40} = [\sin(x\pi/2)]^2$; where $x$ is a measure of the field amplitude and $x = 0.5$ corresponds to the $\pi/2$ pulses optimising the fringe contrast. Operating the fountain for $x \neq 0.5$ leads to a certain reduction of the contrast (and of the short-term stability). It is possible to partially recover the contrast by always applying a full $\pi/2$ pulse at the second passage through the Ramsey cavity [19]. We chose, however, not to vary the field amplitude during the Ramsey time as doing so (e.g. through an operation of a variable microwave attenuator) could have introduced a spurious phase shift and consequently a systematic frequency shift.

The results for the NPL-CsF1 are shown in Fig. 4a where we plotted the collisional frequency shift as a function of $\rho_{40}$. The parameter $\rho_{40}$ was varied in the range 0.18 to 0.82, which corresponds to a microwave field amplitude in the cavity in the range $0.28 < x < 0.72$. The linear function is a very good fit to the experimental points. From the fitted line we find the cancellation point $\rho_{40}^{(C)}$ for the collisional shift. For the experimental conditions of the NPL-CsF1 (asymmetric initial cloud: $\sigma_0^{horizontal} = 0.5$ mm, $\sigma_0^{vertical} = 1.0$ mm and $T_0 = 1.2$ $\mu$K) we got $\rho_{40}^{(C)} = 0.396 \pm 0.012$. Similar results for the PTB-CsF1 are shown in Fig. 4b. The linear function is again a perfect fit to the experimental data. For $T_0 = 2$ $\mu$K (and $\sigma_0 < 1$ mm), the cancellation point is found at $\rho_{40}^{(C)} = 0.298 \pm 0.006$. The same rate of the collisional shift variation was obtained for operation in the vicinity of $3\pi/2$ pulse area (grey data points in Fig. 4b). The quoted uncertainties are statistical one-standard-deviation uncertainties.

To confirm that our observations result from a genuine effect and not from an artefact, e.g. related to variations of the preparing microwave pulse, we used other methods for setting the atom number in the collisional shift measurement, i.e. we varied the detuning of the selection pulse or the MOT loading time. The data obtained with these different methods were consistent within statistical uncertainties. We also repeated the measurements using a new set-up CsF2 being developed at NPL. For $\sigma_0 = 0.7$ mm and $T_0 = 1$ $\mu$K we got $\rho_{40}^{(C)} = 0.43 \pm 0.11$. The values of $\rho_{40}^{(C)}$ obtained in all three experimental set-ups agree well with the theoretical model, although the comparison is limited by uncertainties of $\sigma_0$.

The possibility of the collisional shift cancellation may lead to an improvement of the performance of caesium fountain standards similar to those described above. First we note that in a fountain working at $\rho_{40} = 0.4$ ($x = 0.44$) the fringe contrast (hence the signal to noise ratio and the short-term stability) is only reduced by 3.5% (16% for $\rho_{40} = 0.3$). We further note that, if the launch parameters and the microwave field amplitude in the cavity were sufficiently stable, uncertainty of the collisional shift at parts in $10^{16}$ should be achievable without the need of extrapolation, which would shorten the averaging time. According to the data from Fig. 4, a modest instability of 0.1% in the field amplitude at $\rho_{40}^{(C)}$ would result in a collisional shift not larger than $3\times10^{-17}$. Here, more important would be the stability of the zero-shift point position, which depends on the launch parameters. Alternatively, one can operate a fountain in the vicinity of the zero-shift point and extrapolate the residual shift to zero density. Assuming that the deviation in linearity between $N_{at}$ and the density causes an error in the collisional shift evaluation of less than 10% of the shift itself [18], the uncertainty of the shift may be reduced to parts in $10^{17}$, if one can neglect collisions with atoms residually populating the $|F = 3, m_F \neq 0\rangle$ state. We should mention that the exact position of the zero-shift point may be affected by cavity pulling, an effect that also depends on the number of atoms passing the Ramsey cavity and on the microwave power level in the cavity (thus on $\rho_{40}$) [20]. This effect, however, is estimated to be small ($\leq 10^{-16}$) for the fountains considered in this paper.

The present findings can influence the interpretation of experiments where a fountain clock is run at multiple $\pi/2$ pulse areas to check for power-dependent systematic effects. Typically the pulse area is optimised near a multiple of $\pi/2$ by maximising the contrast of the Ramsey pattern. In general this will lead to a different value of $\rho_{40}$ than for $1\pi/2$ pulse area. As a consequence, the difference in the collisional shift might dominate the frequency shifts observed as a function of pulse area.

To conclude, we have demonstrated experimentally that the frequency shift due to collisions between cold atoms in a caesium fountain clock strongly depends on the proportion of the two clock state populations, $|3,0\rangle$ and $|4,0\rangle$, established after the first passage through the microwave cavity. From the measured values of the shift it follows that collision rates $\lambda_{30}$ and $\lambda_{40}$ have opposite signs, making it possible to cancel the collisional shift for a particular composition of the clock states. This is predicted by theory for the temperatures well below 1 $\mu$K, but not for the typical launch temperatures in the fountains (1-2 $\mu$K). We have shown that the sub-$\mu$K temperatures may be effectively reached during the Ramsey time owing to correlations between position and velocity, which develop in an expanding atomic cloud. The correlations and the reduction of the effective temperature are more pronounced for smaller initial clouds (e.g. when atoms are collected in a MOT, rather than in an optical molasses). In the reported experiments, the collisional frequency shift is cancelled for the fraction of the $|4,0\rangle$ atoms in the range of 30-40%, which can be achieved experimentally with only a minor reduction of the fringe contrast and short-term stability of the fountain frequency standard.

The NPL authors acknowledge discussions with Yuri Ovchinnikov and support from the Time Metrology Programme of the UK National Measurement System. The PTB authors would like to thank Roland Schröder for support in electronics development.

**FIGURE CAPTIONS**

Fig. 1. The theoretical collision rate coefficients as a function of collision energy $E$. The collision energy is given in terms of the Boltzmann constant $k_B$.

Fig. 2. Evolution of the local energy during the ballistic flight. The atoms pass through the Ramsey cavity at $t_{up} = 0.165$ s and $t_{down} = 0.665$ s in the NPL-CsF1 ($t_{up} = 0.130$ s, $t_{down} = 0.695$ s in the PTB-CSF1). The black lines correspond to the initial cloud size of $\sigma_0 = 1$ mm and represent various temperatures at launch (solid line 1 $\mu$K; dashed line 2 $\mu$K; dotted line 5 $\mu$K). The grey solid line corresponds to $\sigma_0 = 5$ mm and initial temperature of 1 $\mu$K.

Fig. 3. Calculated values of the cancellation point for various initial cloud sizes ($\sigma_0$, spherical cloud shapes assumed) and two initial temperatures ($T_0 = 1$ $\mu$K – circles, $T_0 = 2$ $\mu$K – triangles). The points calculated for the same $T_0$ are linked to guide the eye. Note that for $\sigma_0 > 1$ mm ($T_0 \geq 1$ $\mu$K) the cancellation does not occur.
Inset: Calculated collisional frequency shift as a function of the fraction of the $|4,0\rangle$ state population. The position of the zero-shift (cancellation) point $\rho_{40}^{(C)}$ depends on the initial parameters $\sigma_0$ and $T_0$ and the timing $t_{up}$ and $t_{down}$. The slope of the linear function is proportional to the number of atoms in the fountain, with other parameters unchanged.

Fig. 4. Measurement of the collisional frequency shift as a function of the population composition during the Ramsey time. The experimental data are fitted by a linear function. (a) Data from the NPL-CsF1; (b) Data from the PTB-CSF1 (in grey are plotted data-points taken for the fountain operating at $1 < x < 2$).

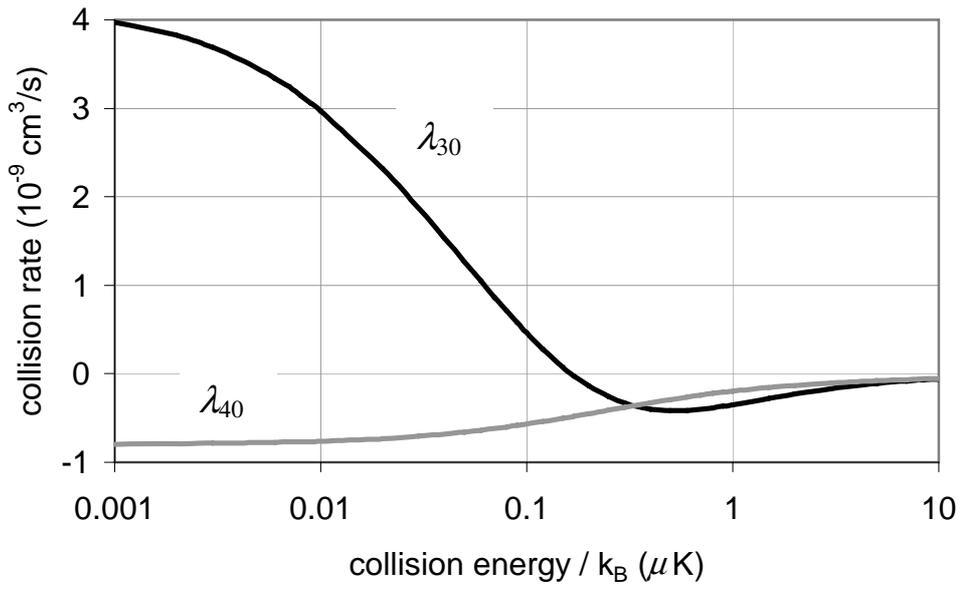

Figure 1.

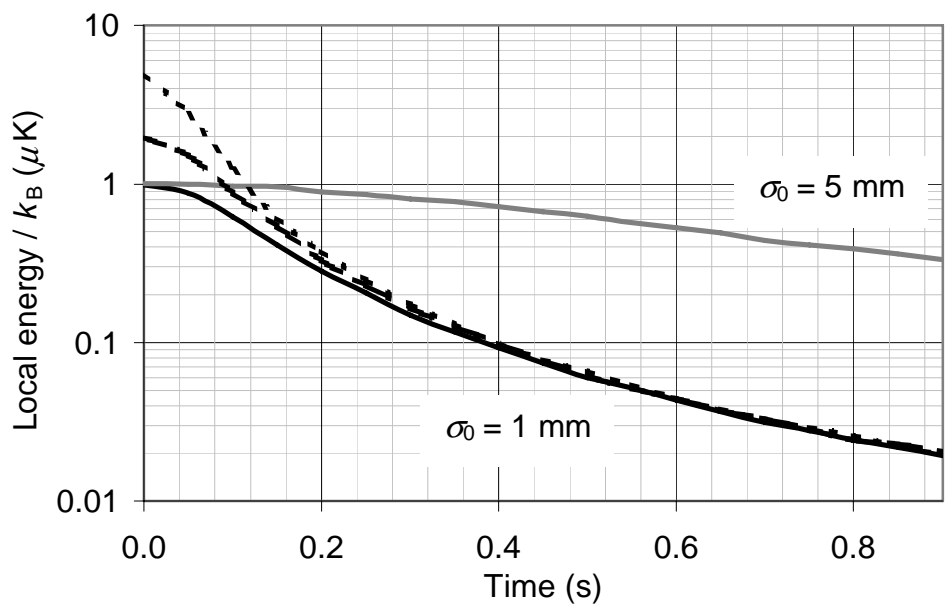

Figure 2.

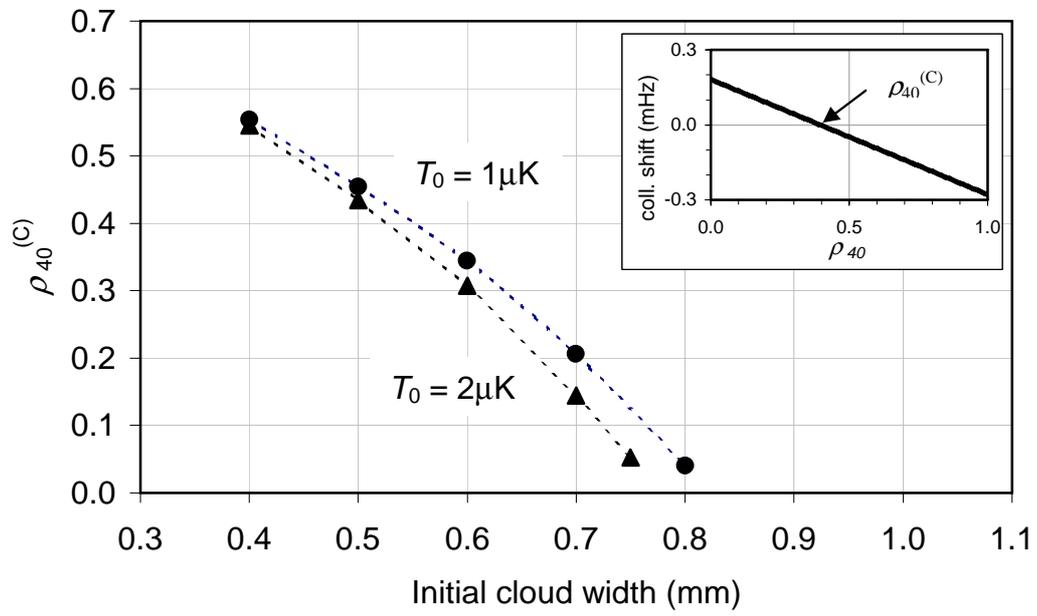

Figure 3.

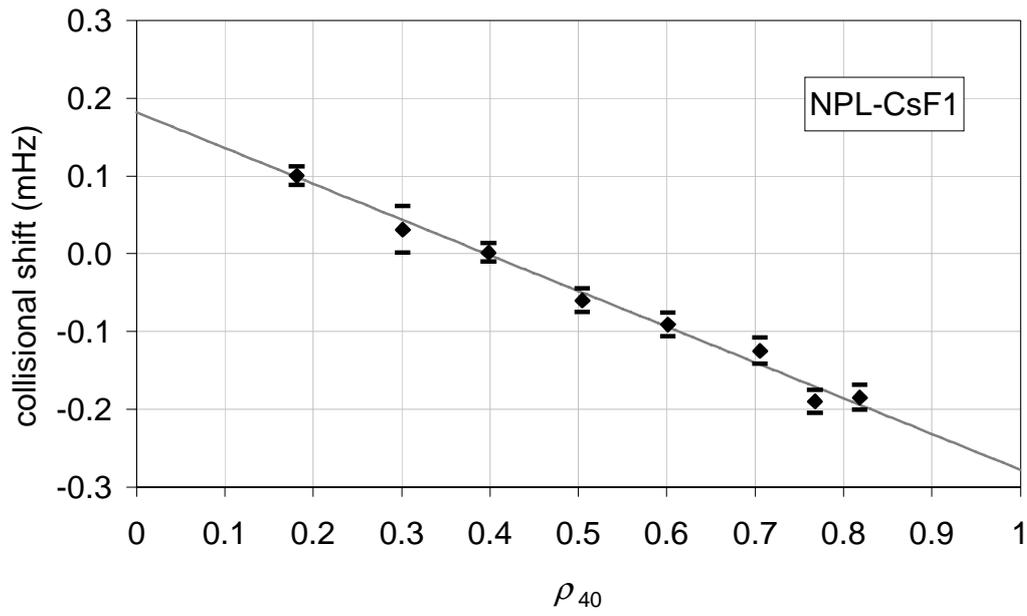

Figure 4 a.

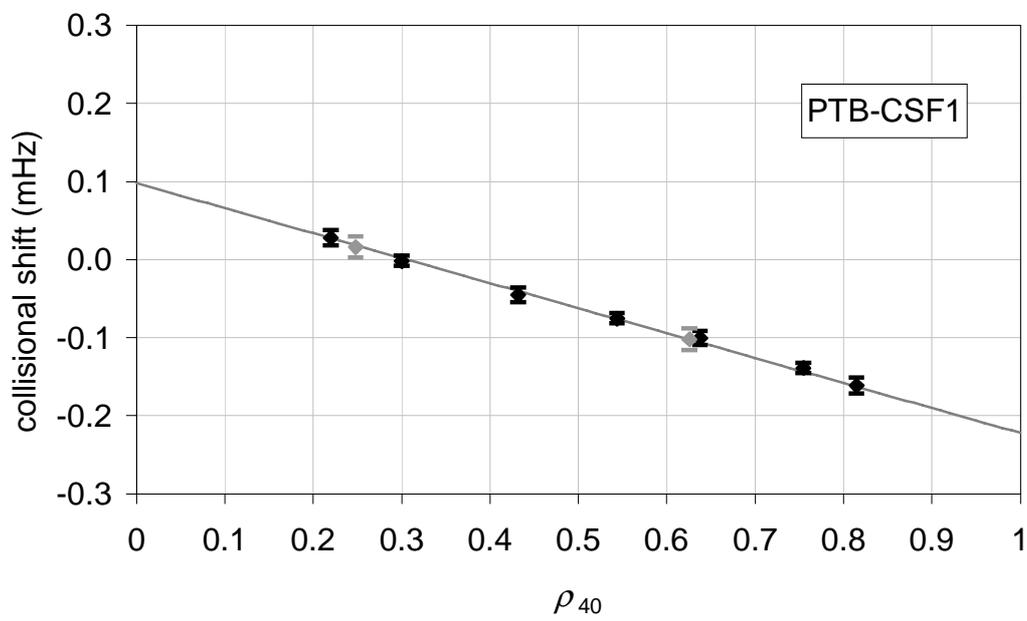

Figure 4 b.